\documentclass[a4paper,11pt]{article}
\pdfoutput=1 

\usepackage{lineno}
\usepackage{mathtools}
\usepackage{siunitx}
\usepackage{jinstpub} 
\usepackage{lineno,hyperref}
\usepackage{xcolor}
\usepackage{verbatim}
\usepackage{amsmath}
\usepackage{float}
\modulolinenumbers[1]

\title{Neutron Imaging Based on Transfer Foil Activation and COTS CMOS Image Sensors}

\author[a,b,1]{M.~P\'erez,\note{Corresponding author.}}
\author[b]{O.~I.~Abbate,}
\author[a,b,c]{J.~Lipovetzky,}
\author[a,c]{F.~Alcalde~Bessia,}
\author[a,b]{F.~A.~S\'anchez,}
\author[a,b,c]{M.~Sofo~Haro,}
\author[a,b]{J.~Longhino,}
\author[a,b,c]{M.~G\'omez~Berisso}
\author[a,c]{and~J.~J.~Blostein}

\affiliation[a]{Centro At\'omico Bariloche, Instituto Balseiro, Universidad Nacional de Cuyo (UNCUYO), Av. E. Bustillo 9500, R8402AGP, San Carlos de Bariloche, R\'io Negro, Argentina}
\affiliation[b]{Comisi\'on Nacional de Energ\'ia At\'omica (CNEA), Av. E. Bustillo 9500, R8402AGP, San Carlos de Bariloche, R\'io Negro, Argentina}
\affiliation[c]{Consejo Nacional de Investigaciones Cient\'ificas y T\'ecnicas (CONICET), Av. E. Bustillo 9500, R8402AGP, San Carlos de Bariloche, R\'io Negro, Argentina}

\emailAdd{martin.perez@ib.edu.ar}

\abstract{In this paper we present a method for obtention of neutron images with Commercial-Off-The-Shelf (COTS) CMOS image sensors through the activation of indium foils. This detection method has been designed specifically for the acquisition of thermal and epitermal neutron images in mixed beams with a high gamma flux, and also for the study of high radioactive samples that are usually placed into research reactor pools. We also present a technique to obtain multi-spectral neutron images taking advantage of the high neutron absorption cross-section of this material in the thermal energy range, as well as around the 1.45\,eV resonance. Measurements performed in a neutron beam of the RA6 nuclear research reactor located in Bariloche, Argentina, confirm the capability of the proposed technique.}


\begin{document}

\maketitle
\flushbottom

\section{Introduction}

Neutron radiography (NR) is a non-destructive imaging technique used in several applications such as non-destructive tests of industrial components, paleontological studies, conservation of art pieces, among others \cite{anderson2009neutron,craft2017applications}. The images are obtained by means of irradiation of the samples with a collimated neutron beam ---usually produced in nuclear research reactors--- and the use of a position sensitive neutron detector in the direct beam for the acquisition of a two-dimensional attenuation map. The traditional detection systems employed in this neutron technique are based on a combination of a scintillator that interacts with neutrons generating visible light, and a 2D image detector to collect the mentioned light and conform the images. Semiconductor detectors ---generally CCD or CMOS cameras--- have replaced traditional detectors like radiographic films due to its high resolution, its ability to perform real-time imaging, and the adjustability of camera settings to experimental conditions. Other 2D neutron detectors are based on \textit{ad hoc} hybrid integrated circuits covered with microchannel plates are also employed to acquire neutron radiographs and microtomographs \cite{tremsin2005efficiency,tremsin2009detection,tremsin2012neutron}. 

NR technique is complementary to conventional X-ray radiography because the attenuation of the incident beam depends on the total neutron cross-section of the sample, which does not depend monotonously on the atomic weight, and besides for some elements allows isotopic discrimination. Hence, by means of NR it is possible to obtain information that is not available by using X-rays \cite{barton2012neutron,craft2016characterization}. 

\begin{figure}[h]
\centering
\includegraphics[width=3.8in]{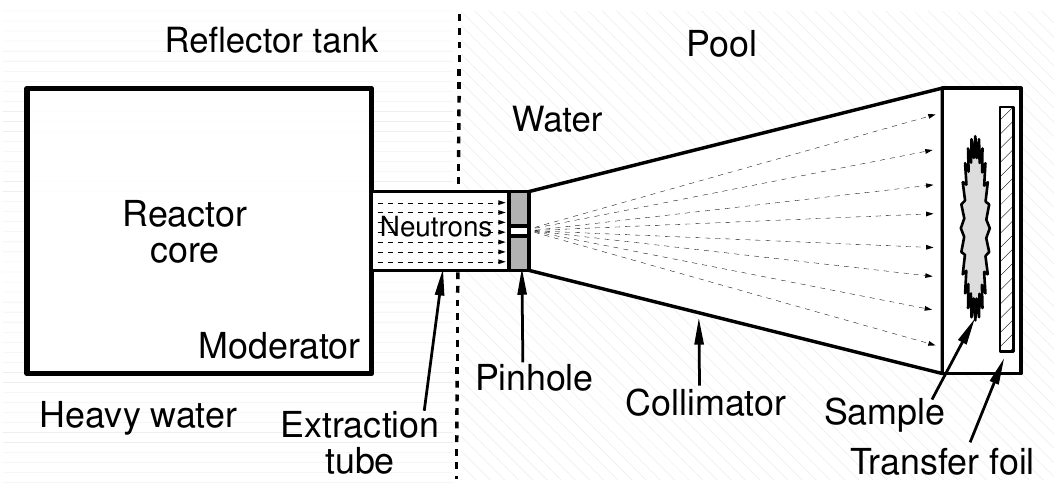}
\caption{Scheme of a typical underwater neutron imaging facility.}
\label{Esquema-facility}
\end{figure}

For non-destructive test of irradiated nuclear fuels NR provide more information than any other non-destructive technique, these kinds of images are used to know the internal condition of the fuel and to study micro-structural properties and different parameters like pellet fragmentation, swelling and cracking, enrichment variations etc. \cite{craft2017conversion,craft2017applications,hayes2016advances,kruvzelova2012neutronographic,richards1982neutron,notea1985gap}. Neutron imaging is also used to perform non-destructive analysis of highly activated structural materials and critical components of nuclear reactors. Due to the risks present in the manipulation of highly-radioactive objects or nuclear fuels, generally it is not possible to use conventional NR facilities\footnote{Generally, these kinds of facilities employ neutron beams with a low gamma component.} to acquire images of them. One of the approaches employed for neutron beam examination of these kinds of samples, is the use of neutron imaging facilities located underwater into pool-type reactors \cite{craft2017applications}. Figure \ref{Esquema-facility} shows a scheme of a typical underwater neutron imaging facility. Thermal and epithermal neutrons are extracted from the neutron moderator ---used to reduce the energy of the fast neutrons generated into the reactor nuclear fuels--- by a tube that connects the reflector tank and the reactor pool. In order to collimate the beam, neutrons have to pass through a pinhole ---placed in the extreme of the tube--- and a collimator. In order to reduce the contribution of fast neutrons present in the beam, the axis of the extraction tube is normally placed tangential to the fuel elements. The walls of these facilities are made of neutron absorbing materials to reduce the flux of not collimated neutrons that could reach the sample.

There are difficulties to place neutron detectors based on semiconductor into reactor pools: On one hand, most of the devices used in NR are sensitive to gamma rays, this hinders their use in underwater facilities because the samples and the reactor core generates high fluxes of photons which can produce a high undesirable counting rate, and a subsequent detector saturation. On the other hand, the high doses of thermal neutrons and gamma photons present in these environments would produce serious damage in semiconductor detectors, reducing considerably its lifespan \cite{bessia2018displacement, Goiffon2009,LeRoch2020}.

Due to the difficulties formerly mentioned, the detection in underwater NR facilities is usually done using an indirect method: A transfer foil (TF) is placed behind the sample ---as shown in Figure \ref{Esquema-facility}---, and exposed to the neutron beam during an interval of time called irradiation time (IT) \cite{craft2017conversion}. Some neutrons, that can cross the sample, reach the TF and are absorbed by it. Then, the "neutron shadow" produced by the sample is thus manifested in the activation level of different TF regions. The sample zones with higher neutron attenuation are traduced in TF zones with lower activity.

After its activation, the TF become temporarily radioactive and decays emitting secondary particles. The levels of activity in the TF increases in the zones where more neutrons were absorbed. In order to obtain the images, the TF is extracted from the pool and scanned using a 2D particle detector \cite{craft2017conversion}.
In previous works, we showed the feasibility of the use of Commercial Off-The-Shelf CMOS image sensors (CIS) in ionizing radiation detection, and thermal neutron detection \cite{perez2016particle, perez2018thermal}. By the employment of this technology, it is possible to obtain high spatial resolution radiation detectors. References \cite{perez2015commercial,perez2016particle,perez2017implementation,haro2020soft,perez2020X-ray,lipovetzky2020multi} shows the response of this type of sensors with different kinds of ionizing particles, it was demonstrated that CIS are sensible to alpha particles, beta particles and also to photons of a wide range of energies. Reference \cite{perez2018thermal} presents the development of a thermal neutron detector based on COTS CMOS sensors covered with a Gd$_2$O$_3$ conversion layer, in this case the events detected by the semiconductor were produced by conversion electrons generated in the conversion layer after the interaction with neutrons. In reference \cite{perez2021high} we presented a position-sensitive neutron detection technique based on a CIS covered with nanoparticles of NaGdF$_4$, and we prove that the intrinsic spatial resolution of this on-line detection method is better than (15$\pm$6)\,$\mu$m.

In this work, we present a method for the acquisition of neutron images based on the activation of indium transfer foils and its subsequent activity recording with a CIS. We also present a method for the acquisition of multi-spectral images based on the use of In filters to modify the neutron spectra employed for the activation of the transfer foils.
Next section presents the readout electronics employed for the detection of the beta particles emitted from the In decays, the characteristics of the indium transfer foils, and the details of the proposed method. In Section \ref{Irradiations}, we will present the details of the experiments performed at the RA6 nuclear research reactor. Section \ref{transmissions-probability-interaction} shows a theoretical estimation of the neutron attenuation in the employed samples and an analysis to estimate the neutron absorption probability in the TF. In Section \ref{Neutron-image} we will show the first neutron images. Section \ref{Multispectral-neutron-image} presents the method for the obtention of a multi-spectral neutron image of the samples.  
Finally, Section \ref{Conclusions} presents the conclusions of this work.

\section{Materials and Methods}

In this section we introduce the different components used to perform the measurements presented in this work, as well as the details of the proposed method.

\subsection{The CMOS sensors}

The integrated circuit used in this work is the APTINA MT9M001, which is a 1/2-inch CMOS monochrome active-pixel image sensor composed by 1280\,$\times$\,1024 pixels with 5.2\,$\mu$m\,$\times$\,5.2\,$\mu$m pitch \cite{Onsemi}. The sensor active area is covered with approximately 3.7\,$\mu$m of metal layers ---embedded in an inter-metal dielectric---, and a layer of approximately 3.8\,$\mu$m of polymers which form the micro-lenses of each pixel \cite{bessia2018displacement}. 

The readout electronics employed for the measurements presented in this work is composed by two boards: a ``Camera Module'' which contains a socket where the CMOS sensor is positioned, and the ``Arducam Camera Shield'' that uses an FPGA to read the sensor and transfer the data to a computer via USB \cite{Arducam}. A Python script is used to collect the data from the USB as well as to set different parameters of the sensor, for example: gain, integration time, white balance, among others. During the experiments the sensor was configured to obtain the maximum gain of the in-pixel amplifiers, and the maximum integration time. All the measurements were done acquiring 0.5 frames per second.

\subsection{The In transfer foils}
The materials employed as transfer foil for this application must have four main characteristics: Firstly, they must have a high neutron absorption cross-section. Secondly, the radioactive nuclei produced by the neutron absorption have to decay emitting charged particles that must be detectable with the employed image sensor. In third place, the activated nuclei produced by neutrons in the TF must have a half-life 
large enough in order to have time to remove it from the reactor and place it over the sensor surface with measurable activity level. Finally, the half life of the transfer foils must be short enough to allow the measurement and quantification of a detectable activity during acquisition times that not exceed a few tens of minutes \cite{domanus1987neutron,firestone1996table}. Besides, it is necessary a TF with a decay time short enough to allow its re-utilization after some hours without a considerable previous activity. Indium is the most widely used material for the fabrication of transfer foils because it meets all the above-mentioned requirements. The isotope of $^{115}$In has a natural abundance of 95.71\,\%, and its most probable interaction with 
thermal and epithermal neutrons produces the following nuclear reaction:

\begin{equation}
 n+^{115}In\,\rightarrow\,^{116m_1}In + \gamma.
\label{activacion} 
\end{equation}

\begin{figure}[h]
\centering
\includegraphics[width=4in]{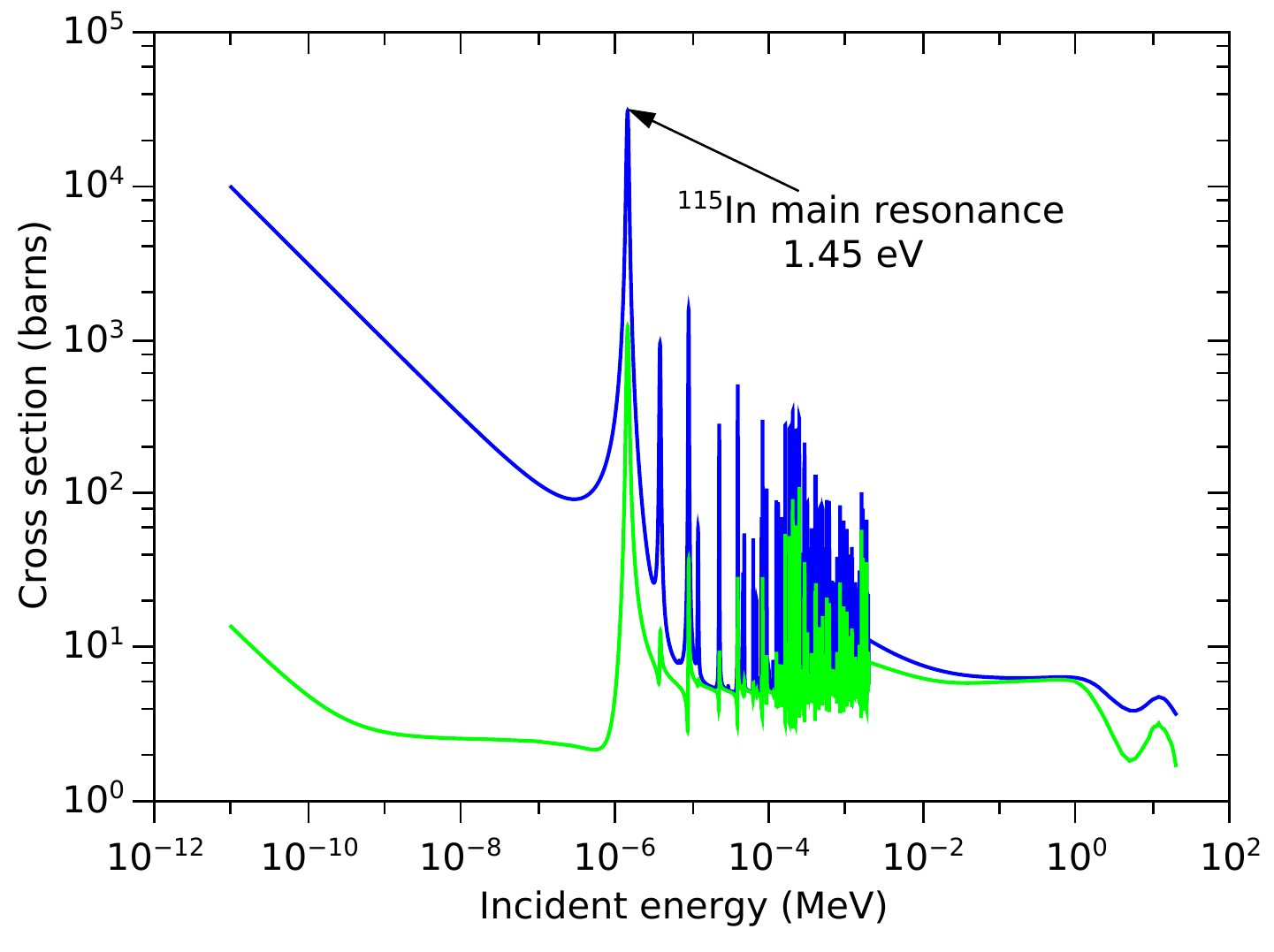}
\caption{Elastic scattering cross-section ---green--- and total cross-section ---blue--- for the isotope of $^{115}$In \cite{chadwick2006endf}.}
\label{cross-section}
\end{figure}

Figure \ref{cross-section} shows the elastic scattering cross-section ---green curve--- and the total cross-section ---blue curve--- for the $^{115}$In. 
This isotope, has a neutron resonance at 1.45\,eV that produces most of the absorptions in the epithermal range. In the thermal range the material has a typical $1/\upsilon$ absorption behavior due to its relatively very low scattering cross-section. The massive product generated by the Reaction \ref{activacion} is the $^{116m_1}$In, which is the first metastable state of  $^{116}$In that has a half life of 54.29 minutes, and decays by $\beta^-$ to $^{116}$Sn. As a result of the decay a $\beta^-$ particle and a neutrino are emitted, each particle carried a fraction of the total energy produced in the reaction with a continuous spectrum.  
In the case of the $\beta^-$ particles emitted in the $^{116m_1}$In decay, the energy distribution has a peak of occurrence at approximately 150\,keV and a maximum energy of 1.014\,MeV \cite{IAEANuclearData,firestone1996table,chadwick2006endf}.

\subsection{The proposed method}

To perform the measurements presented in this work, we used indium foils with a thickness of approximately 88\,$\mu$m. The In foils were cut and adhered to bases specially designed for this purpose, these bases are made on Al and provide a mechanical support for the TF. Aluminum is a low-activation material and has low neutron scattering cross-section, thus after the irradiations the signal produced by Al during the acquisition time is negligible.

\begin{figure}[!t]
\centering
\includegraphics[width=4in]{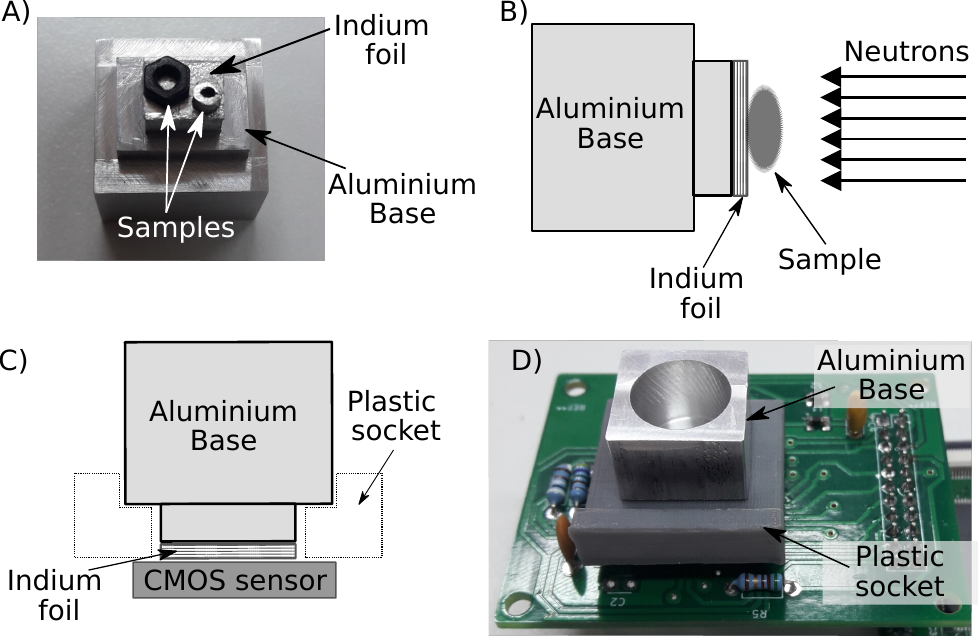}
\caption{a) Image of the samples over the In foil, and the aluminum base. b) Scheme of the experimental setup employed during the In foil activation. c) Scheme of the experimental setup employed to detect the $\beta^-$ particles emitted by the In foil. d) Image of the aluminum base mounted over the CMOS sensor.}
\label{Diagramas-method}
\end{figure}

The samples were placed over a set of TF+base (Figure \ref{Diagramas-method} ~(a)) and fastened with aluminum tape. To perform the TF neutron activation, the whole set was irradiated as shown in the scheme of Figure \ref{Diagramas-method}~(b). In the covered regions of the TF the detected neutrons have to pass through the sample to reach the indium surface. The contribution of neutrons that reach the TF from its back side is very low due to the low neutron scattering produced in the aluminum. After the activation, the samples were removed from the base, and the TF was placed over the CMOS sensors surface as shown in the scheme of the Figure \ref{Diagramas-method}~(c).  A plastic socket -- designed for this application-- is used for positioning the TF over the active area of the sensor (Figure \ref{Diagramas-method}~(d)). 
Then, the CMOS image sensor was employed to read the $\beta^-$ particles through the acquisition of images at a rate of 0.5\,FPS. Finally, the acquired images were processed using a software specially developed for this application. The script analyzes all the recorded images and add the charge produced in each pixel, thus generating the resulting image.

\section{Irradiations at the RA6 Nuclear Research Reactor}
\label{Irradiations}

\begin{figure}[!t]
\centering
\includegraphics[width=4in]{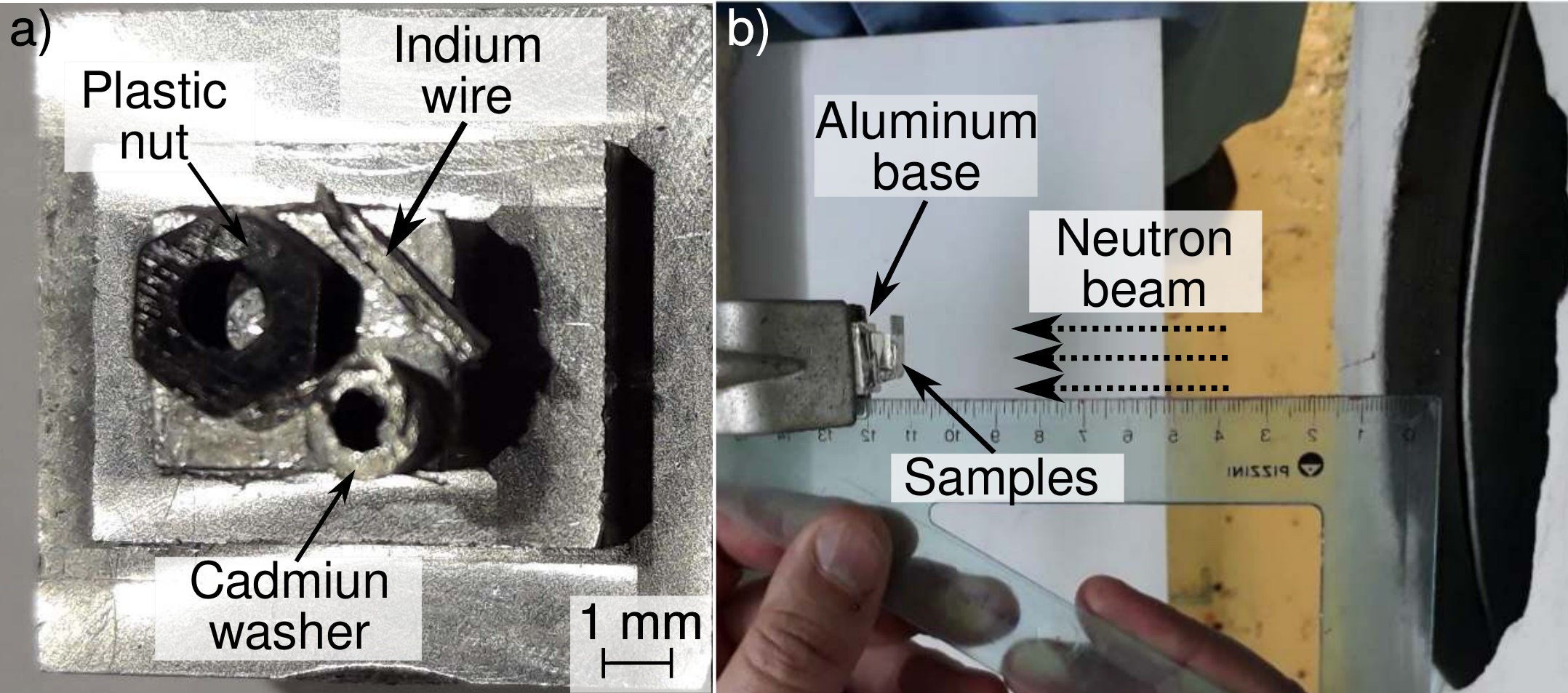}
\caption{a) Samples used to perform the neutron radiographs placed over the indium TF and the aluminum base: a plastic nut, a cadmium washer and an indium wire. b) Experimental setup of the irradiations performed in the BNCT Facility of the RA6 Reactor.}
\label{Diagramas-method-In-BNCT}
\end{figure}

In order to analyze the feasibility of the proposed detection method, we performed two irradiations in the neutron beam present in the BNCT facility of the RA6 nuclear research reactor, located at the Bariloche Atomic Centre, Argentina. 
This beam was selected for these irradiations because it has a mixed thermal-epithermal energy spectrum with a flux\footnote{This neutron flux was measured in the beam exit port which is a hole with a circular section of 15 cm diameter employed to extract the neutrons from the reactor core. The beam exit port is located in the shield (made of lead and borated polyethylene) of the installation.} of $\sim$ 3.5$\times$10$^8$\,n/cm$^2$s, which is the highest of the RA6 reactor. Although the neutron beam in the BNCT facility is not collimated, the neutron flux in this installation allows to achieve high activity levels, which implies a large amount number of events in the obtained images. On the other hand, this neutron beam has an ephitermal component which allows to modify the neutron spectra by filters of different materials that produces attenuation at different neutron energy ranges.
Figure\,\ref{Diagramas-method-In-BNCT}~(a) shows the samples placed over the transfer foil, in these experiments we acquired images of a plastic nut, a cadmium washer, and an indium wire. The mentioned samples have a thickness of 2\,mm, 1.7\,mm, and 1\,mm respectively. Figure\,\ref{Diagramas-method-In-BNCT}~(b) shows the experimental setup employed during the irradiations. 

\begin{figure}[!h]
\centering
\includegraphics[width=4in]{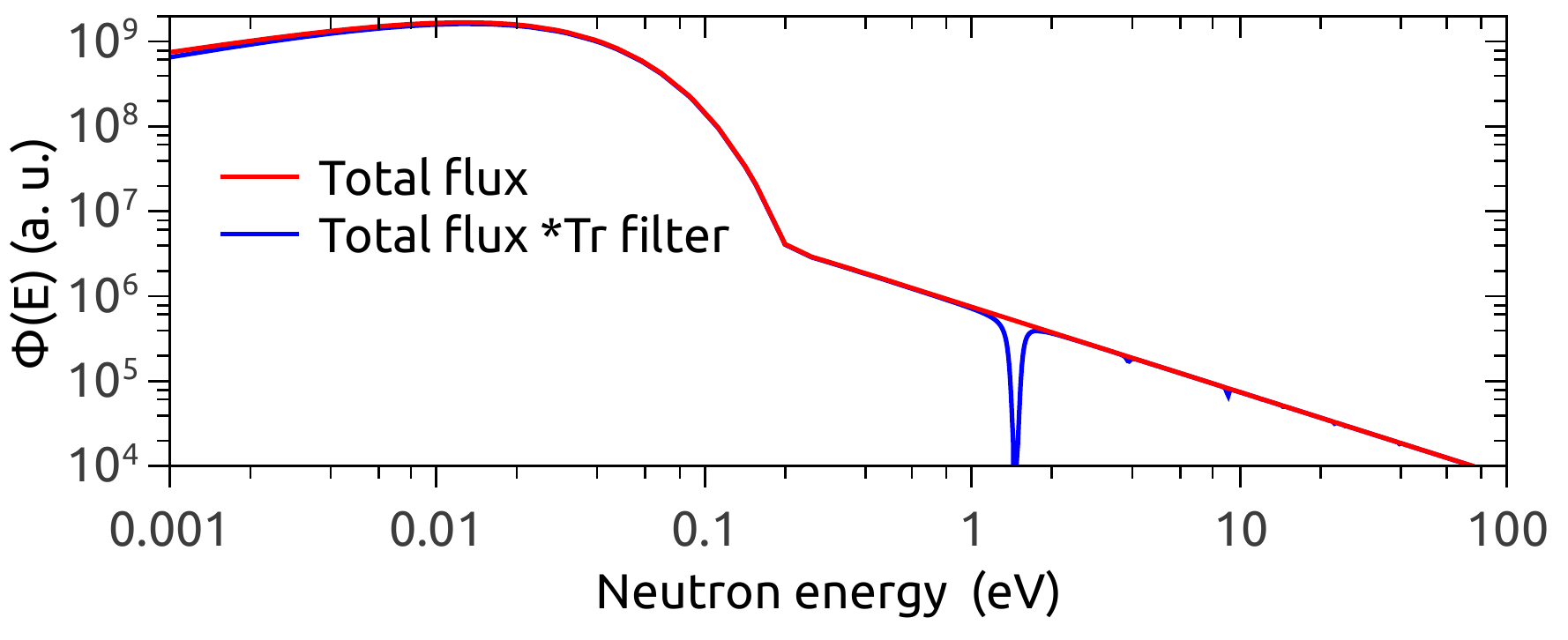}
\caption{Red curve: estimation of the neutron spectrum in the BNCT beam. Blue curve: estimation of the neutron spectrum of the BNCT beam weighted by the indium filter transfer.}
\label{Flux-BNCT}
\end{figure}

We activated the transfer foil at a reactor power of 1\,MW with two different configurations: In a first irradiation the TF was activated with the entire incident spectrum composed by thermal and epithermal neutrons, the red curve of Figure\,\ref{Flux-BNCT} shows an estimation of this spectrum. The samples were placed at a distance of (116$\pm$2)\,mm from the beam exit port. During the irradiation, we measured the flux received by the TF with the standard neutron activation method. For this purpose, we placed two Au-Cu wires with 1.55\% of Au at both sides of the samples.
In order to obtain the value of the epithermal flux, one wire was covered with a cadmium sheath. Under this experimental conditions, the TF was activated with a thermal flux of (8.8$\pm$0.6$\times10^7$)\,n/cm$^2$s and an epithermal flux of (1.3$\pm$0.1$\times10^7$)\,n/cm$^2$s.

During a second irradiation, we interposed a (37.0$\pm$0.5)\,$\mu$m thick sheet of indium between the beam exit port and the samples in order to absorb neutrons with energies around the main resonance of In at 1.45\,eV. In this configuration the TF was activated with a neutron energy spectrum that is similar to the presented in the blue curve of Figure\,\ref{Flux-BNCT}.

\section{Discussion}

In this section, we will present the theoretical calculations of the neutron transmissions in the In filter employed to modify the incident neutron beam, in the formerly mentioned samples, as well as an estimation of the neutron absorption probability in the TF as functions of the incident neutron energy.
We also calculated the neutron absorption probability in the TF. Finally, we will present the neutron images acquired after the irradiations, and a multi-spectral neutron image of the mentioned samples.

\subsection{Neutron transmissions and absorption probability in the TF}
\label{transmissions-probability-interaction}

As it is shown in \cite{chadwick2011endf}, the neutron transmission in any material can be calculated with the following equation:

\begin{equation}
\label{eq-transmision}
    Tr(E)=e^{-nx\sigma_{Tot}(E)}\,,
\end{equation}
where:

$x$ = thickness of the object,

$n$ = density of number of atoms, 

$\sigma_{Tot}(E)$ = total cross-section \cite{chadwick2011endf}.

\begin{figure}[h]
\centering
\includegraphics[width=4in]{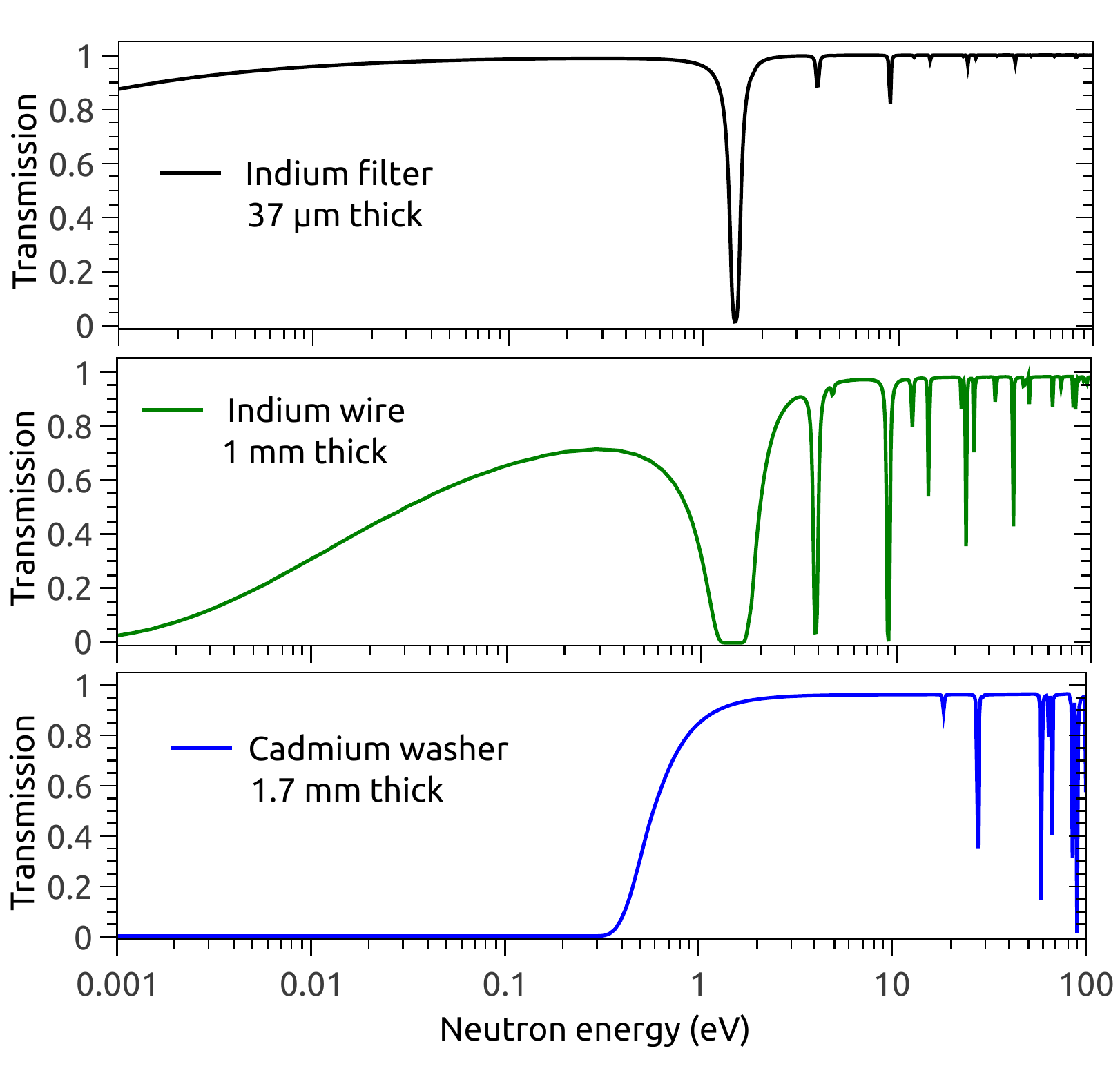}
\caption{Transmission as a function of the incident neutron energy for the In filter ---upper panel---, the In wire used as a sample ---central panel---, and for the Cd washer ---lower panel---.}
\label{transmission}
\end{figure}

Figure\,\ref{transmission} shows the one-dimensional transmission curves ($Tr$) as a function of the incident neutron energy ($E$) obtained by Equation \ref{eq-transmision} for the In filter, the Cd washer, and the In wire used as sample. It is possible to observe that the In filter transmission is greater than 0.9 for the entire range of energies presented, except around the absorption resonance at 1.45\,eV where $Tr$ is almost zero.
In the case of the In wire used as a sample, the transmission in the thermal range is less than 0.8 and is zero around the resonances of 1.45, 3.8 and 9.1\,eV. The thickness of the In wire is greater than the In foil used as a filter, for this reason its transmission in the thermal range is more influenced by the $1/\upsilon$ behavior of the cross-section. In addition, in Figure\,\ref{transmission} it can also be observed that the neutron transmission in the Cd washer used as a sample is zero for neutrons with energies of less than 0.3\,eV and is approximately 0.93 for energies around 1.45\,eV.

\begin{figure}[h]
\centering
\includegraphics[width=4.1in]{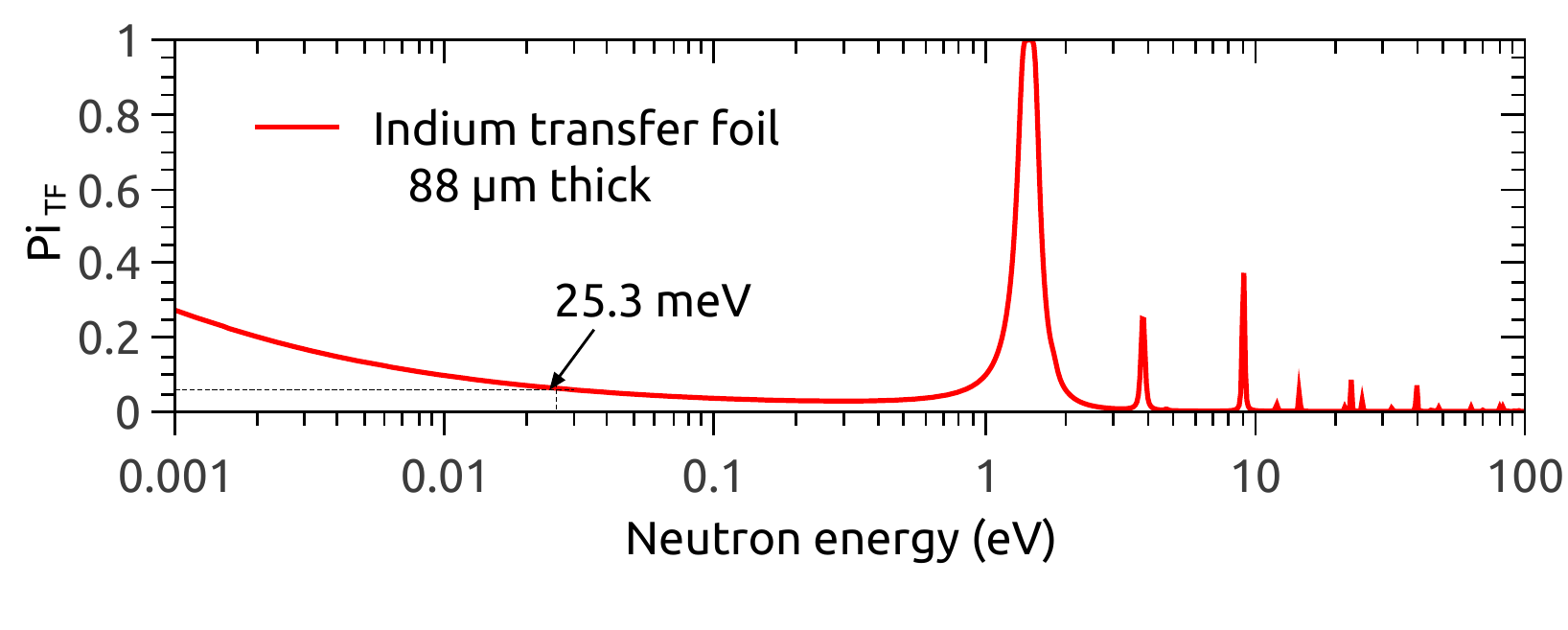}
\caption{Probability of interaction in the transfer foil as a function of the incident neutron energy.}
\label{Prob-inter-TF}
\end{figure}

The curve presented in Figure\,\ref{Prob-inter-TF} shows the interaction probability in the TF ($Pi_{TF}$) as a function of the incident neutron energy, and was calculated with the following equation\footnote{This approximation is valid because the scattering cross-section is negligible with respect to the absorption cross-section.}:
\begin{equation}
    Pi_{TF}=1- Tr_{TF}\,,
\end{equation}
\noindent where $Tr_{TF}$ is the transmission in the TF calculated with Equation\,\ref{eq-transmision}. In Figure\,\ref{Prob-inter-TF} it is possible to observe that for energies around the resonance of 1.45\,eV  the $Pi_{TF}$ is 100\%, this is the energy range where the TF has the maximum neutron absorption, and therefore highest sensitivity. On the other hand, in Figure  \,\ref{Prob-inter-TF} can also be observed that the $Pi_{TF}$ is approximately 6\% for thermal neutrons of 25.3\,meV.

\subsection{Neutron images}
\label{Neutron-image}

The intensity recorded in each pixel of the images presented in this section are proportional to the reaction rate ($R$). The microscopic reaction rate ($R$) can be calculated with the following equation \cite{de1972neutron}:

\begin{equation}
\label{ritmo-de-reacciones}
    R=\frac{\Sigma^*}{k(1-e^{-\lambda t_i}) e^{-\lambda t_d} (1-e^{-\lambda t_m})}\,,
\end{equation}

where:

$\Sigma^*$= Number of measured counts,

$\lambda$= decay constant,

$t_i$= irradiation time,

$t_d$= decay time,

$t_m$= measurement time.

\noindent The Equation\,\ref{ritmo-de-reacciones} expresses the microscopic reaction rate per target nucleus. Assuming that the self-shielding of neutrons in the transfer foil is negligible, the constant $k$ of the Equation\,\ref{ritmo-de-reacciones} can be calculated with the following expression:

\begin{equation}
    k=\frac{C \, y \, \epsilon \, \eta\,m\,N_A}{M \lambda}\,,
\end{equation}
where:

$C$ = concentration,

$y$ = particle emission yield,

$\epsilon$ = detector efficiency,

$\eta$ = isotopic abundance,

$m$ = sample mass,

$N_A$= Avogadro number,

$M$ = element molar mass.

It is worth to notice that the constant $k$ only depends on the intrinsic properties of the In transfer foil, and the detection efficiency of the CIS. Besides, $k$ is independent of the irradiation time, decay time, and acquisition time. Therefore, if the attenuation of the $\beta$ particles in the TF is neglected, we can consider that $k$ is constant for all the measurements. Since the detection efficiency is unknown, $R$ values will be reported in arbitrary units, assuming $k$=1.

\begin{figure}[h]
\centering
\includegraphics[width=3.9in]{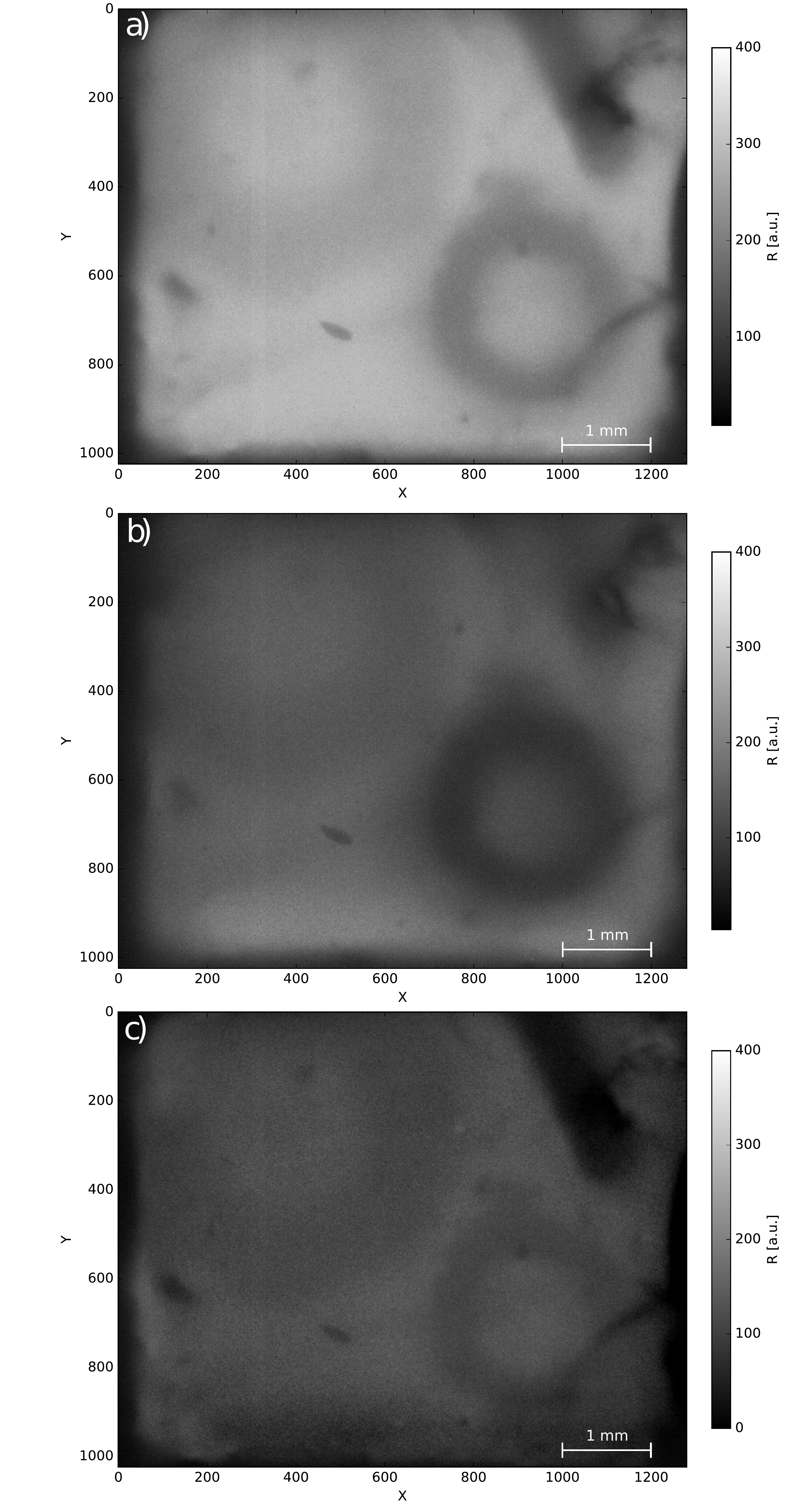}
\caption{a) Reaction rate $R_a$ obtained with the entire neutron spectrum of the BNCT beam. In the image it is possible to observe the "neutron shadows" of the indium wire ---upper right corner---, the cadmium washer ---lower right corner---, and the plastic nut ---upper left corner---, see photograph in Figure\,\ref{Diagramas-method-In-BNCT}~(a). b) Reaction rate $R_b$ obtained using an In filter to modify the incident neutron spectrum. c) Reaction rate $R_c$ obtained by calculating the difference $R_a - (R_b/Tr_{ filter})$. On the three images the reactions rates are expressed in the same arbitrary units and with the same gray scale.}
\label{Neutrografias}
\end{figure}

In the first experiment performed in the BNCT facility, the entire set ---base+TF+samples--- was irradiated with the full incident neutron spectrum\footnote{Without any filter between the beam exit port and the samples.} during 125 minutes. Then, the TF was read during two periods of acquisition of 49 and 56 minutes, after decays times of 37 and 109 minutes respectively. Figure\,\ref{Neutrografias}~(a) shows the obtained reaction rate ($R_a$), where it is possible to observe the "neutron shadows" of the indium wire ---upper right corner---, cadmium washer ---lower right corner---, and plastic nut which can be seen more tenuous in the upper left corner. In the borders of Figure\,\ref{Neutrografias}~(a), as well as in certain areas of them\footnote{The dark zones can be observed below the In wire shadow, at the right of the Cd washer shadow, and also below the plastic washer shadow.}, it is possible to observe dark zones which are produced by the lack of uniformity and depressions of the transfer foil. In these zones the TF is not in contact with the sensor surface, causing that many of the beta particles emitted during the decays do not reach the sensor active volume. These defects were produced during the assembly of the TF on the aluminum base due to the extreme ductility of In.  

Figure\,\ref{Neutrografias}~(b) shows the reaction rate ($R_b$) obtained in the second irradiation ---in which we interposed the indium filter in the incident beam---, $R_b$ is presented in the same arbitrary units of $R_a$. The samples were irradiated for 121 minutes and the TF was read during a time interval of 30 min, after a decay time of 30 min. It can be observed that the shadow of the cadmium washer is darker than that obtained in the first irradiation ---Figure\,\ref{Neutrografias}~(a)---, because in this case the thermal component is preponderant.
In Figure\,\ref{Neutrografias}~(b) it also can be observed that the shadow produced by the plastic nut cannot be seen clearly ---because the full neutron flux was attenuated by the In filter---, and that the shadow produced by the In wire is almost imperceptible because most of the neutrons with energies around the resonance of 1.45\,eV ---which is the part of the spectrum where the In wire produces the greatest contrast--- were also attenuated by the In filter.

As can be seen in Figure\,\ref{transmission}, the transmission of the In filter ($Tr_{filter}$) is almost constant in the thermal energy range. In order to cancel the contribution to the image due to thermal neutrons, we generated a third image by calculating the difference between the reaction rate $R_a$ and the reaction rate $R_b$ weighted by the filter transmission in the thermal range:
\begin{equation}
    R_c=R_a-\frac{R_b}{Tr_{filter}}.
\end{equation}
Replacing in Equation\,\ref{ritmo-de-reacciones} we have:
\begin{equation}
\label{ritmo_de_reaccion_c}
    R_c=\frac{\Sigma^*_a}{(1-e^{-\lambda t_{ia}}) e^{-\lambda t_{da}} (1-e^{-\lambda t_{ma}})}
    -\frac{\Sigma^*_b}{(1-e^{-\lambda t_{ib}}) e^{-\lambda t_{db}} (1-e^{-\lambda t_{mb}})\cdot Tr_{filter}}.
\end{equation}

\noindent After the cancelation of thermal contribution, the reaction rate $R_c$ is produced mainly by neutrons around the epithermal resonances, and is specially dominated by the main resonance of 1.45\,eV. Figure\,\ref{Neutrografias}~(c) shows the result obtained after the application of Equation\,\ref{ritmo_de_reaccion_c} to the images presented in Figures\,\ref{Neutrografias}~(a) and (b).
The "neutron shadow" of the In wire shown in Figure\,\ref{Neutrografias}~(c) have more contrast than that observed in Figure\,\ref{Neutrografias}~(a) and (b), this is so because the reaction rate $R_c$ is mainly dominated by neutrons with energies around the indium resonance, where the In wire transmission is almost zero. On the other hand, the Cd washer and the plastic nut produce a small attenuation for neutrons around 1.45\,eV, therefore in Figure\,\ref{Neutrografias}~(c) it is possible to observe that “neutron shadows” of these samples are fainter than that produced by the In wire.

NIST provides libraries of the Continuous Slowing Down Approximation Ranges (CSDA) of electrons \cite{berger1992estar}. For beta particles of 150\,keV, which is the most probable energy emitted in the $^{116m_1}$In decay \cite{IAEANuclearData}, the CSDA range in Si is $\sim$152\,$\mu$m. This result allow estimating the order of magnitude of the intrinsic spatial resolution of the implemented method.

\subsection{Multi-spectral neutron image}
\label{Multispectral-neutron-image}

\begin{figure}[h!]
\centering
\includegraphics[width=3in]{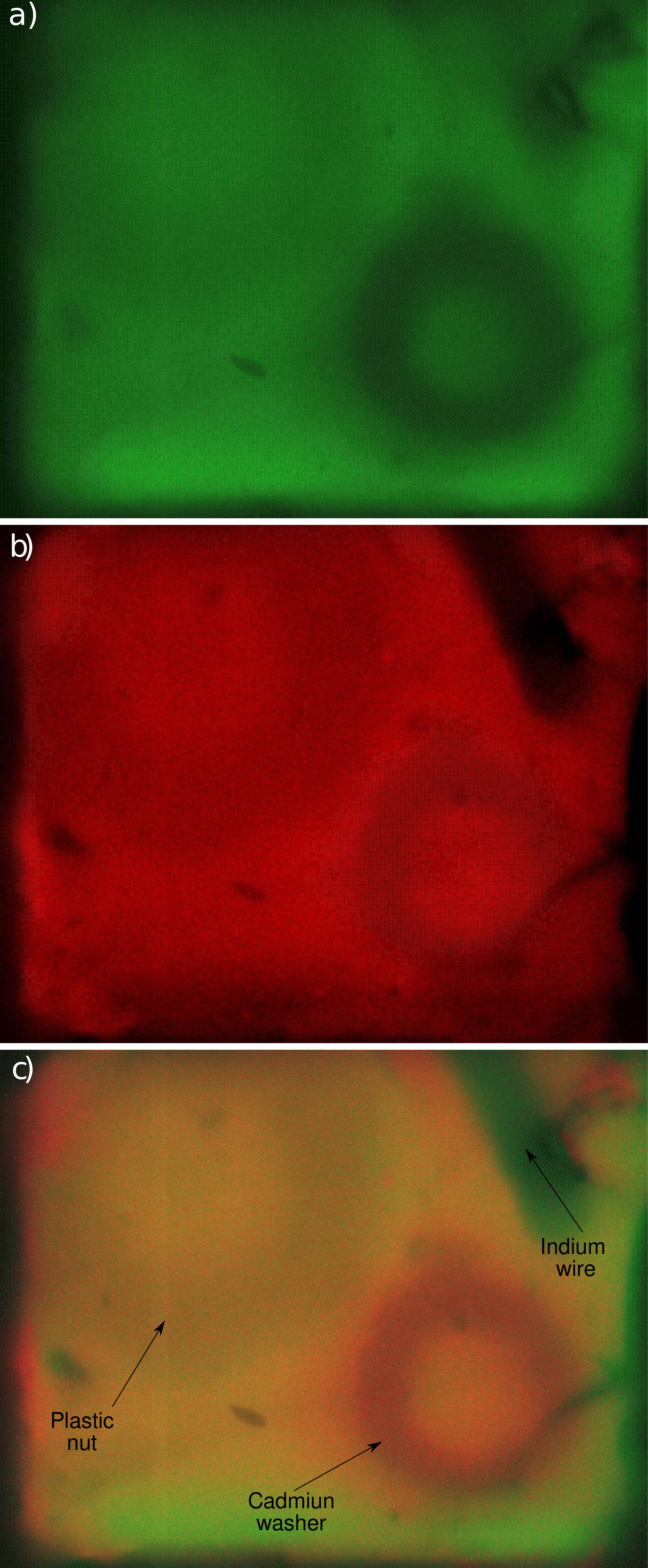}
\caption{a) Neutron image presented in Figure\,\ref{Neutrografias}~(b) represented in green color, this image was obtained mainly with thermal reactions. b) Neutron image presented in Figure\,\ref{Neutrografias}~(c) represented in red color, this image was obtained with reactions mainly around the indium resonance. c) multi-spectral image generated by adding the images (a) and (b). The green color indicates that the sample produces more attenuation around the 1.45 eV resonance, while the red color indicates that the sample produces more attenuation in the thermal range.}
\label{Multiespectral}
\end{figure}

Figure\,\ref{Multiespectral}~(c) shows a multi-spectral image generated with the neutron radiographs presented in Figure\,\ref{Neutrografias}~(b) ---thermal component--- and Figure\,\ref{Neutrografias}~(c) ---epithermal component---. In this case, the reaction rate of the image ($R_{multi-spectral}$) was obtained with the following equation:
\begin{equation}
    R_{multi-spectral}= R_{th} <green> +R_{1.45 eV} <red>,
\end{equation}
where:
\begin{itemize}
    \item $R_{th}$ = $R_b$ = reactions mainly produced by thermal neutrons represented in green color in Figure\,\ref{Multiespectral}~(a).
    \item $R_{1.45\,eV}$ = $R_c$ = $R_a - \frac{R_b}{Tr_{filter}}$ = reactions produced by neutrons with energies around the resonance of 1.45\,eV represented in red as shown in Figure\,\ref{Multiespectral}~(b).
\end{itemize}

This multi-spectral image provides more information than a grayscale neutron image, because its color indicates the energy range in which the samples are more absorbent.
In Figure\,\ref{Multiespectral}~(c) the green areas are originated when the reactions produced by thermal neutrons ($R_{th}$) are greater than that produced around the resonance of 1.45\,eV ($R_{1.45\,eV}$). This situation occurs at zones where the sample absorbs neutrons around the resonance, which is observed in the shadow produced by the In wire.
Reciprocally, in the red areas $R_{1.45\,eV}$ is greater than $R_{th}$, this occurs when the sample produces a greater attenuation in the thermal range, for example in the  Cd washer shadow. Finally, it is possible to  find intermediate tonalities between pure green and pure red according to the proportion of the components $R_{1.45\,eV}$ and $R_{th}$ in different image regions.

\section{Conclusions}
\label{Conclusions}

We have experimentally demonstrated that it is possible to obtain neutron images by the activation of indium transfer foils and its subsequent off-line beta activity reading with Commercial-Off-The-Shelf CMOS image sensors. For this test, we employed the neutron beam available at the BNCT facility, which was selected because it has the highest neutron flux available at the RA6 nuclear research reactor, nevertheless this is a not collimated beam. 
For this preliminary test, we have prioritized the employment of a high neutron flux instead of the beam collimation, thus increasing the signal to noise ratio. For further applications, it will be possible to improve the spatial resolution of the acquired neutron images by using a neutron beam with a better collimation (higher L/D).

By means of the use of an indium filter in the incident beam, we have shown that it is possible to acquire multi-spectral neutron images using the first indium resonance in the epithermal range, and its high neutron cross-section in the thermal range. 
These multi-spectral images provide more information than the traditional neutron images because the observed color allows identifying the sample absorption energy ranges.

Using a set of transfer foils and filters of different materials with neutron absorptions resonances at other energies —for example Au, Mn, Cu— it will be possible to obtain more information of the samples analyzing the absorption in other ranges of energy. Besides, a pair of filter foil and transfer sheet of the same material will be specially sensible for the detection of this material into the sample. 
The technique presented in this work is specially suitable for the acquisition of high spatial resolution multi-spectral neutron images  of samples with high levels of activity ---for example nuclear fuel and structural components of nuclear research reactors--- employing "in-pile" neutron image facilities.
For this application, the intense gamma background in such environments would make the use of on-line neutron detectors impracticable. Furthermore, the high activity level of these samples would difficult its relocation out of the reactor pool.

\section{Acknowledgments}

This work was supported by ANPCyT (Argentina) under project PICT 2018-2886, and by SIIP Universidad Nacional de Cuyo (Argentina) under project Cod. 06/553. The authors would like to thanks to Ignacio Artola for his technical assistance, and to the members of the RA6 staff Julio Marín, Fabricio Brollo and Soia Capararo.

\bibliography{bibliografia}

\end{document}